\documentclass [prl,twocolumn,superscriptaddress]{revtex4-1}

\usepackage{graphicx}
\usepackage{subfigure}
\usepackage{amsmath}
\usepackage{mathrsfs}
\usepackage{amsthm}
\usepackage{amssymb}
\usepackage{hyperref}
\usepackage{xcolor}
\usepackage{csquotes}

\hypersetup{colorlinks=true, linkcolor=black, citecolor=black, urlcolor=blue}
\allowdisplaybreaks

\newcommand{\ket}[1]{\left| #1 \right>} 
\newcommand{\bra}[1]{\left< #1 \right|} 

\begin{document}

\title{Cavity-quantum-electrodynamical toolbox for quantum magnetism}
 
\author{Farokh Mivehvar}
\email[Corresponding author: ]{farokh.mivehvar@uibk.ac.at}
\affiliation{Institut f\"ur Theoretische Physik, Universit{\"a}t Innsbruck, A-6020~Innsbruck, Austria}
\author{Helmut Ritsch}
\affiliation{Institut f\"ur Theoretische Physik, Universit{\"a}t Innsbruck, A-6020~Innsbruck, Austria}
\author{Francesco Piazza}
\affiliation{Max-Planck-Institut f\"{u}r Physik komplexer Systeme, D-01187 Dresden, Germany}

\begin{abstract}
The recent experimental observation of spinor self-ordering of ultracold atoms in 
optical resonators has set the stage for the exploration of emergent
magnetic orders in quantum-gas--cavity systems. Based on this platform, we introduce 
a generic scheme for the implementation of long-range quantum spin Hamiltonians
composed of various types of couplings, including Heisenberg and 
Dzyaloshinskii-Moriya interactions. Our model is comprised of an effective two-component 
Bose-Einstein condensate, driven by two classical pump lasers and coupled to a single 
dynamic mode of a linear cavity in a double $\Lambda$ scheme. 
Cavity photons mediate the long-range spin-spin interactions with spatially 
modulated coupling coefficients, where the latter ones can be tuned by modifying spatial profiles of the pump lasers.
As experimentally relevant examples, we demonstrate that by properly choosing the spatial profiles of the pump lasers 
achiral domain-wall antiferromagnetic and chiral
spin-spiral orders emerge beyond critical laser strengths. 
The transition between these two phases can be observed in a
single experimental setup by tuning the reflectivity of a mirror. We also
discuss extensions of our scheme for the
implementation of other classes of spin Hamiltonians.
\end{abstract}

\maketitle
\emph{Introduction.}---Quantum magnetism
plays a crucial role in many phenomena in condensed matter physics~\cite{Sachdev2008quantum-magnetism}, 
including for instance high-temperature
superconductivity~\cite{Anderson2004highT-supercondoc} and
spin liquids~\cite{Balents2010spin-liquid}. In
materials, there exist different forms of interactions between electronic spins.
The Heisenberg interaction, originating from the
isotropic quantum exchange interaction between electrons, favors ferromagnetic (FM) or
antiferromagnetic (AFM) ordering~\cite{Parkinson2010quantum-spin-model}. 
The more exotic Dzyaloshinskii-Moriya (DM) 
interaction~\cite{Dzyaloshinsky1958, Moriya1960a, Moriya1960b},
stemming from a relativistic antisymmetric exchange interaction,
favors chiral states such as spin spiral (SS) and 
skyrmion~\cite{Braun2012-nanomagnetism, Nagaosa2013skyrmion-rev, Fert2013skyrmion-rev,
Hellman2017RevModPhys-interface-magnetism,Uchida2006helical-order, Bode2007chiral-inversion-symm, 
Yoshida2012conical-spin-spiral, Haze2017chiral-spin},
with potential applications in
spintronics~\cite{Baltz2018AF-spintronic}. 

That said, it is not an easy task to modify
the strength and nature of spin-spin interactions in materials,
putting stringent constraints on controlled experimental
explorations as well as technological applications.
Therefore, the notion of simulating quantum magnetism
using more controllable systems has emerged~\cite{Micheli2006-QM-toolbox-molecules, Cole2012-BH-SOC, Radi2012-QM-SOC, Cai2012-SOC-BH, Gong2015-DM-SOC}, with experimental implementations using ultracold
atoms~\cite{Trotzky2008exchange-interc, Simon2011AF, Struck2011frustrated-class-magnetism, 
Greif2013quantum-magnetism, Struck2013XY-Ising,Hart2015Hubbard-AF, Greif2015AF-correl, 
Mazurenko2017Fermi-Hubbard-AF,tolra_dipolar_2013,gross_2015,biedermann_2015,browaeys_2016,gross_2016},
molecules~\cite{Ye_molecules_2013}, and ions~\cite{roos_ions_2014,monroe_ions_2014}.

Another promising route for simulating quantum magnetism with atoms is to utilize 
cavity-mediated long-range spin-spin interactions, which do not require extremely low
temperatures to come into play~\cite{Strack2011spin-glass,
Gopalakrishnan2011glassiness, Buchhold2013-Dicke-spin-glass,Mivehvar2017disorder,Norcia2018spin-exchange,Davis2018spin-ex}.
These interactions exist in all regimes, including very deep Mott phase 
where the atomic tunnelling between lattice sites is completely suppressed.
This is a remarkable feature of cavity-mediated long-range spin interactions, 
similar to dipolar interactions between polar molecules~\cite{Ye_molecules_2013}. 
The first step in this direction has been taken very recently by the experimental
realization of density and spin self-ordering with ultracold bosonic atoms
inside an optical cavity~\cite{Landini2018spin-texture, Kroeze2018spin-texture}, 
and with thermal atoms near a retroreflecting mirror~\cite{Krei2018thermal-spin-text}. 
These experiments have basically realized a long-range
Heisenberg model, with an emergent domain-wall AFM order.

\begin{figure}[b!]
\centering
\includegraphics [width=0.46\textwidth]{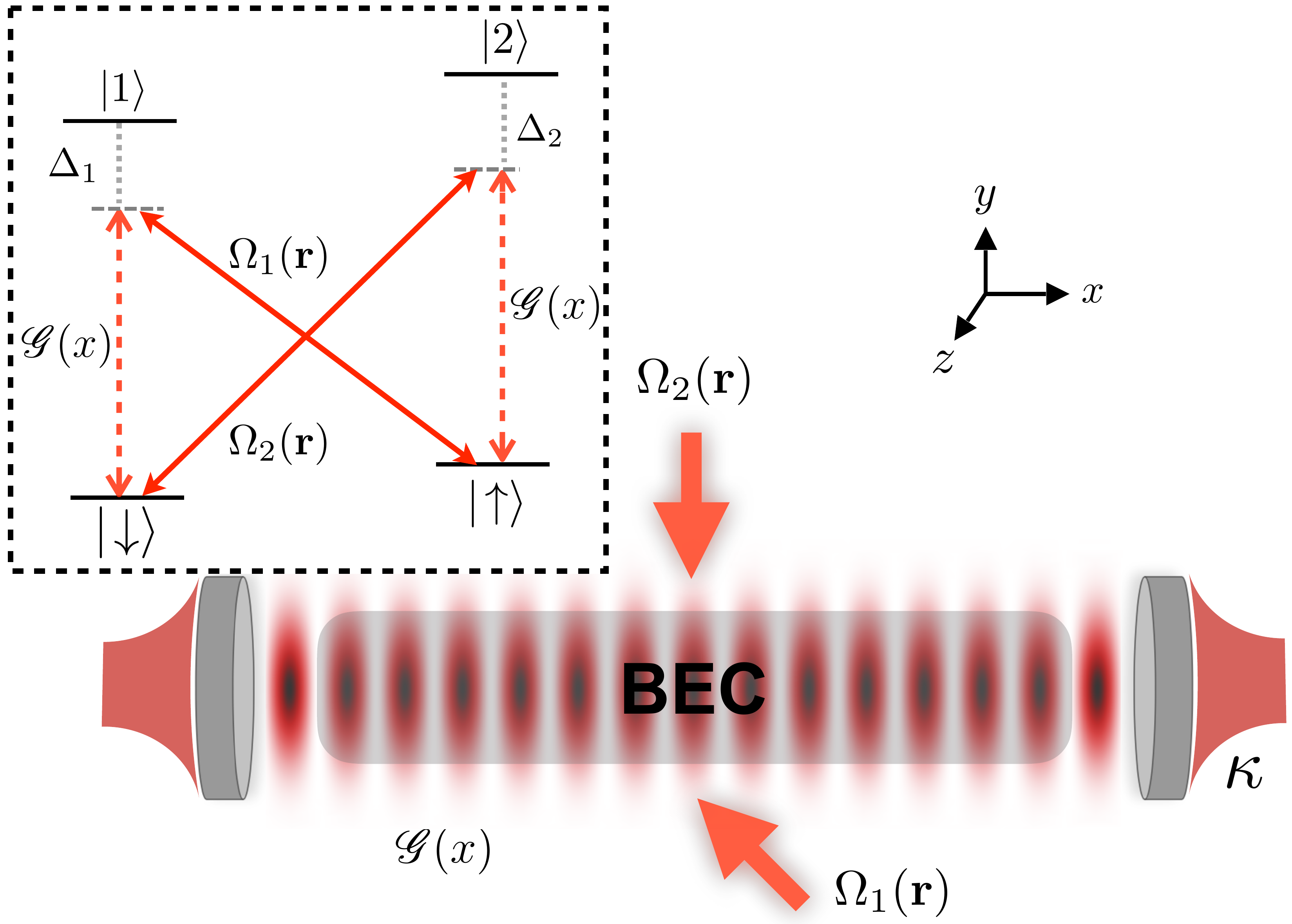}
\caption{Schematic view of a transversely pumped multi-component BEC in 2D inside a cavity. 
The inset depicts the internal atom-photon coupling scheme in a double $\Lambda$ configuration.} 
\label{fig:lin-cavity-geom--Lambda-scheme}
\end{figure}

Motivated by the recent experimental progress, 
in this Letter we demonstrate how to engineer a variety of spin models in the framework of cavity QED. 
We consider a multi-component Bose-Einstein condensate (BEC), 
which is driven by two
pump lasers and coupled to a single dynamic
mode of a standing-wave cavity in a double $\Lambda$ scheme  as depicted in 
Fig.~\ref{fig:lin-cavity-geom--Lambda-scheme}.
By adiabatic elimination of the atomic excited states and the cavity field,
we derive an effective long-range spin-$1/2$ Hamiltonian 
with spatially modulated coupling coefficients. The spatial modulations arise from the
interference among different electromagnetic modes, and thus depend
crucially on the spatial profiles of the two pump lasers and the cavity-mode function.

\begin{figure}[t!]
\centering
\includegraphics [width=0.485\textwidth]{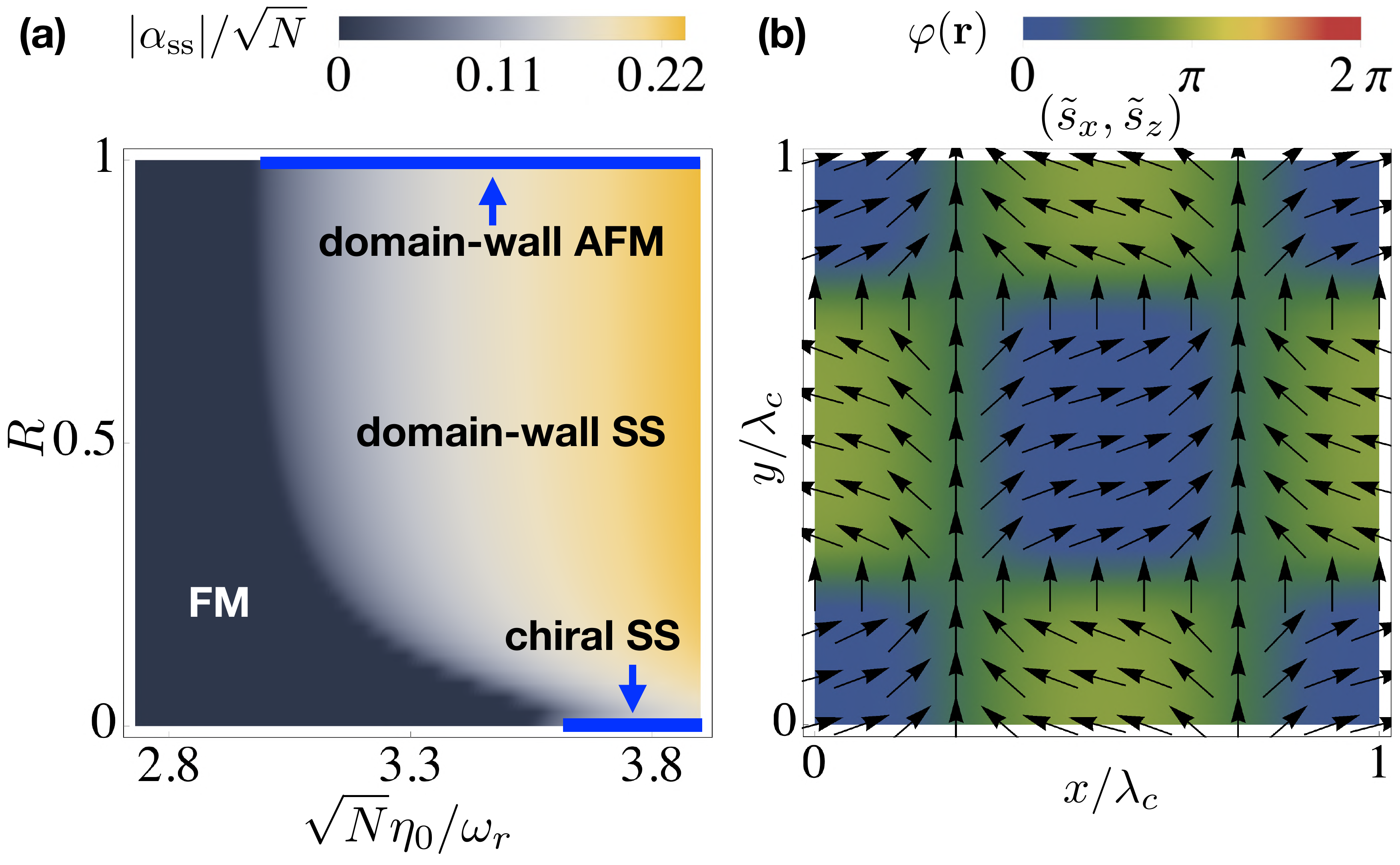}
\caption{The phase diagram of the system (a) and domain-wall AFM spin texture (b). 
(a) The phase diagram of the system is shown in the $\{R,\sqrt{N}\eta_0/\omega_r\}$ plane, 
where the color code indicates the rescaled cavity field amplitude $|\alpha_{\rm ss}|/\sqrt{N}$.  
(b) The projection of the normalized local spin $\tilde{\mathbf s}(\mathbf{r})$ on 
the $\tilde{s}_x$-$\tilde{s}_z$ plane is displayed as a function of $\mathbf{r}$ 
for two standing-wave pump lasers along the $y$ direction corresponding to $R=1$. 
The color code indicates
the spin angle with respect to the $\tilde{s}_x$ axis, 
$\varphi=\tan^{-1}(\tilde{s}_z/\tilde{s}_x)$. Inside each domain,
all spins are oriented with an angle $\varphi_{\rm D}\simeq0.081\pi$ or $\pi-\varphi_{\rm D}$,
while in boundaries they rotate quite rapidly. The parameters are set to
$(\Delta_c,\kappa,N\mathscr{G}_0^2/\Delta_{1,2},\delta)=(-10,5,-1,-0.1)\omega_r$ 
and $\sqrt{N}\eta_0=3.8\omega_r$
with the self-consistent field amplitude $|\alpha_{\rm ss}|/\sqrt{N}\simeq0.22$.
Here $N$ is the total particle number and $\omega_r\equiv\hbar k_c^2/2M$.} 
\label{fig:domain_wall_ST}
\end{figure}

\begin{figure}[t!]
\centering
\includegraphics [width=0.49\textwidth]{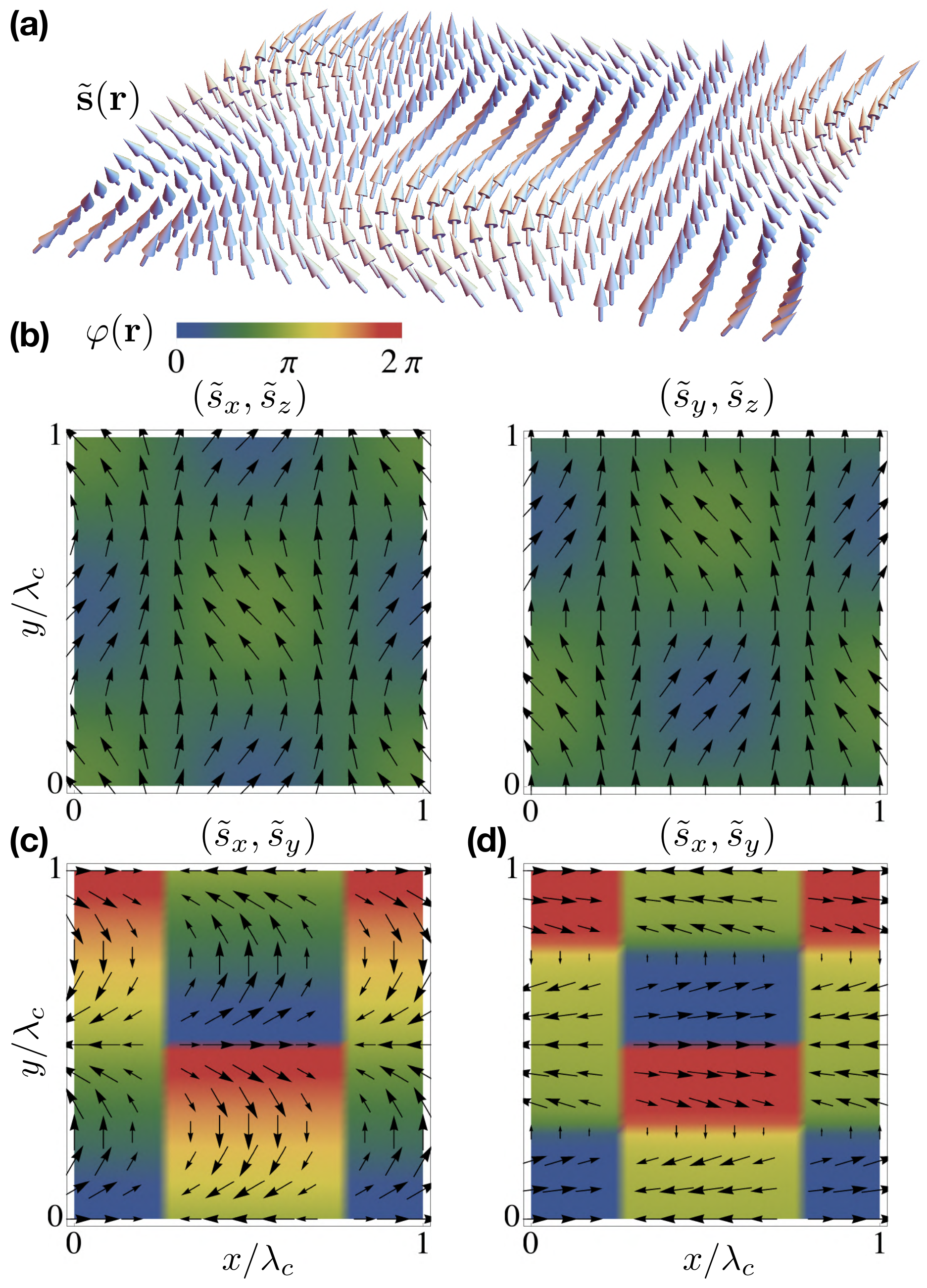}
\caption{Emergent SS texture. 
(a) The normalized local spin $\tilde{\mathbf s}(\mathbf{r})$ is shown
for two counterpropagating pump lasers along the $y$ direction corresponding to $R=0$.
The spin exhibits a transverse conical SS texture with magnetic domains.
The latter is most clearly evident in the projections of $\tilde{\mathbf s}(\mathbf{r})$
in the $\tilde{s}_x$-$\tilde{s}_z$and $\tilde{s}_y$-$\tilde{s}_z$ planes as shown in panel (b).
The spin does a full $2\pi$ rotation in the $\tilde{s}_x$-$\tilde{s}_y$ plane
along the $y$ direction over one $\lambda_c$, as can be seen in panel (c).
The spin projection in the $\tilde{s}_x$-$\tilde{s}_y$ plane for $R=0.2$
is illustrated in the panel (d) with pronounced magnetic domains.
The color code in each
figure indicates the respective spin angle, e.g., $\varphi=\tan^{-1}(\tilde{s}_y/\tilde{s}_x)$ 
in the panels (c,d). The parameters are 
the same as Fig.~\ref{fig:domain_wall_ST},
with the self-consistent field amplitude $|\alpha_{\rm ss}|/\sqrt{N}\simeq0.14$ and $0.21$ for 
$R=0$ and $0.2$, respectively.} 
\label{fig:chiral_ST}
\end{figure}

We show that the resultant quantum spin models are not restricted to only 
Heisenberg-type interactions as theoretically~\cite{Strack2011spin-glass, Gopalakrishnan2011glassiness, 
Buchhold2013-Dicke-spin-glass, Mivehvar2017disorder} and 
experimentally~\cite{Norcia2018spin-exchange, Davis2018spin-ex, Landini2018spin-texture, Kroeze2018spin-texture} 
explored thus far, but can also include DM-type interactions.
For two standing-wave pump lasers in the transverse direction
as the experiment of Ref.~\cite{Kroeze2018spin-texture}, 
the spin Hamiltonian reduces to a long-range Heisenberg Hamiltonian with a spatially
periodic coupling constant. The photon-mediated Heisenberg-type interaction 
induces an achiral, domain-wall AFM spin order beyond a
critical laser strength; see Fig.~\ref{fig:domain_wall_ST}(b). 
On the other hand for two counterpropagating running-wave
pump lasers in the transverse direction, the spin Hamiltonian also
includes DM-type interactions. As a result, the steady state exhibits a chiral, conical SS 
order beyond a critical laser strength as shown in Fig.~\ref{fig:chiral_ST}(a)-(c).
At the end we also discuss extensions of our scheme to implement more
general spin models, paving the way for the realization of
exotic magnetic orders and frustrated states in atom-cavity systems.

\emph{Model.}---Consider four-level ultracold bosonic atoms trapped in 
two dimensions (2D) in the $x$-$y$ plane inside a linear cavity
and illuminated in this plane by two classical pump lasers 
as depicted in Fig.~\ref{fig:lin-cavity-geom--Lambda-scheme}. 
The transition $\uparrow\>\leftrightarrow1$ ($\downarrow\>\leftrightarrow2$) is driven by the first (second) pump 
laser with the position-dependent Rabi coupling $\Omega_1(\mathbf{r})$ [$\Omega_2(\mathbf{r})$],
where $\mathbf{r}=(x,y)$. On the other hand, the transitions 
$\downarrow\>\leftrightarrow 1$ and $\uparrow\>\leftrightarrow 2$ are
coupled to a single (initially empty) 
cavity mode with the same coupling strength $\mathscr{G}(x)=\mathscr{G}_0\cos(k_cx)$.
In this double $\Lambda$ configuration, $\tau=\{\downarrow,\uparrow\}$ 
are the relevant (e.g., Zeeman or hyperfine pseudo) spin states 
and $\{1,2\}$ are some auxiliary electronic excited states, 
with energies $\{\hbar\omega_\downarrow=0, \hbar\omega_\uparrow,\hbar\omega_{1},\hbar\omega_{2}\}$. 
For the moment we do not specify the spatial profiles of the Rabi
couplings $\Omega_{1,2}(\mathbf{r})$; specific examples will be given later.
However, their frequencies $\{\omega_{\rm p1},\omega_{\rm p2}\}$ are assumed to be in close resonance
with the cavity frequency $\omega_c=ck_c$. The pump and cavity frequencies 
are all far red detuned from the atomic transition frequencies
in that $\Delta_1\equiv(\omega_{\rm p1}+\omega_{\rm p2})/2-\omega_1$ and
$\Delta_2\equiv\omega_{\rm p2}-\omega_2$ are large.
Nonetheless, two-photon Raman transitions are close to resonant:
$\omega_c-\omega_{\rm p1}\approx\omega_{\rm p2}-\omega_c\approx\omega_\uparrow$.

After adiabatic elimination of the atomic excited states for large $\Delta_{1,2}$~\cite{SM-spin-texture2018},
the system is described by the many-body Hamiltonian 
$\hat{H}=-\hbar\Delta_c\hat{a}^\dag\hat{a}+\int \hat\Psi^\dag(\mathbf{r})\hat{\mathcal H}_0\hat\Psi(\mathbf{r})d\mathbf{r}$,
where $\Delta_c\equiv(\omega_{\rm p1}+\omega_{\rm p2})/2-\omega_c$, and 
$\hat{a}$ and $\hat\Psi=(\hat\psi_\uparrow,\hat\psi_\downarrow)^\top$ are
the photonic and two-component atomic bosonic annihilation field operators, respectively.
The single-particle Hamiltonian density reads,
\begin{align} \label{eq:single-particle-H-den}
\hat{\mathcal H}_0=
\begin{pmatrix}
\frac{\mathbf{p}^2}{2M}+\hbar\delta+\hat{V}_\uparrow(\mathbf{r}) & \hbar\hat\Omega_{\rm R}(\mathbf{r}) \\
\hbar\hat\Omega^\dag_{\rm R}(\mathbf{r}) & \frac{\mathbf{p}^2}{2M}+\hat{V}_\downarrow(\mathbf{r})
\end{pmatrix},
\end{align}
with the potential operators 
$\hat{V}_\uparrow(\mathbf{r})=\hbar\hat{a}^\dag\hat{a}|\mathscr{G}(x)|^2/\Delta_2+\hbar|\Omega_1(\mathbf{r})|^2/\Delta_1$,
$\hat{V}_\downarrow(\mathbf{r})=\hbar\hat{a}^\dag\hat{a}|\mathscr{G}(x)|^2/\Delta_1+\hbar|\Omega_2(\mathbf{r})|^2/\Delta_2$,
and the two-photon Raman-Rabi coupling operator 
$\hat{\Omega}_{\rm R}(\mathbf{r})=\Omega_1^*(\mathbf{r})\hat{a}\mathscr{G}(x)/\Delta_1
+\hat{a}^\dag\mathscr{G}^*(x)\Omega_2(\mathbf{r})/\Delta_2$. Here $M$ is the atomic mass, 
$\mathbf{p}=(p_x,p_y)$ is the atomic momentum operator, and 
$\delta\equiv\omega_\uparrow-(\omega_{\rm p2}-\omega_{\rm p1})/2+B_{\rm ext}$ with $B_{\rm ext}$
being an external magnetic field to tune the internal atomic levels. The origin of the potentials and the Raman-Rabi coupling
can be intuitive understood based on photon scattering processes. For instance, $\hat{V}_\uparrow$ 
is a Stark shift due to absorbing a cavity and/or first-pump photon and re-emitting it to the same mode
by spin-up atoms. 

\emph{Effective long-range spin-spin interactions.}---In the Born-Oppenheimer (or adiabatic)
limit corresponding to large $\Delta_c$ and/or large cavity-field decay rate $\kappa$~\cite{Ritsch2013RMP}, 
the cavity-field operator reaches in a short time scale its steady state $i\hbar\partial_t \hat{a}_{\rm ss}=[\hat{a}_{\rm ss},\hat{H}]=0$,
\begin{align} \label{eq:ss-a}
\hat{a}_{\rm ss}=\frac{\int \mathscr{G}^*(x)\left[\frac{1}{\Delta_1}\Omega_1(\mathbf{r})\hat{s}_-(\mathbf{r})
+\frac{1}{\Delta_2}\Omega_2(\mathbf{r})\hat{s}_+(\mathbf{r})\right]d\mathbf{r}}
{\Delta_c+i\kappa-\int |\mathscr{G}(x)|^2\left[\frac{1}{\Delta_1}\hat{n}_\downarrow(\mathbf{r})
+\frac{1}{\Delta_2}\hat{n}_\uparrow(\mathbf{r})\right]d\mathbf{r}},
\end{align}
where $\hat{n}_\tau(\mathbf{r})=\hat{\psi}^\dag_\tau(\mathbf{r})\hat{\psi}_\tau(\mathbf{r})$ and
$\hat{s}_+(\mathbf{r})=\hat{s}_-^\dag(\mathbf{r})
=\hat{\psi}^\dag_\uparrow(\mathbf{r})\hat{\psi}_\downarrow(\mathbf{r})$. 
The cavity field
can be thus integrated out by formally
substituting the steady-state photonic field operator~\eqref{eq:ss-a} in the Hamiltonian $\hat{H}$,
yielding an effective atom-only Hamiltonian. This effective atomic Hamiltonian
consists of a (local) single-particle part for the center-of-mass motion, 
plus a long-range interaction part for the spin degree of freedom~\cite{SM-spin-texture2018}
\begin{align}\label{eq:eff-spin-H}
\hat{H}_{\rm spin}&=\int \Big\{
\sum_{\beta=x,y}J_{\rm Heis}^{\beta}(\mathbf{r}',\mathbf{r})\hat{s}_\beta(\mathbf{r}')\hat{s}_\beta(\mathbf{r})
\nonumber\\
&+J_{\rm DM}^{z}(\mathbf{r}',\mathbf{r})
\left[\hat{s}_x(\mathbf{r}')\hat{s}_y(\mathbf{r})-\hat{s}_y(\mathbf{r}')\hat{s}_x(\mathbf{r})\right]\nonumber\\
&+J_{\rm c}^{xy}(\mathbf{r}',\mathbf{r})
\left[\hat{s}_x(\mathbf{r}')\hat{s}_y(\mathbf{r})+\hat{s}_y(\mathbf{r}')\hat{s}_x(\mathbf{r})\right]
\Big\}d\mathbf{r}d\mathbf{r}'\nonumber\\
&+\int B_z(\mathbf{r})\hat{s}_z(\mathbf{r}) d\mathbf{r},
\end{align}
where $\hat{\mathbf s}(\mathbf{r})=\hat\Psi^\dag(\mathbf{r})\pmb\sigma\hat\Psi(\mathbf{r})$ is the local pseudospin operator
($\pmb\sigma$ is the vector of Pauli matrices).
The first line in Eq.~\eqref{eq:eff-spin-H} corresponds to the $x$ and
$y$ components of a Heisenberg-type interaction $\hat{\mathbf s}(\mathbf{r}')\cdot\hat{\mathbf s}(\mathbf{r})$.
The second line corresponds to the $z$ component of a 
DM-type interaction $\hat{\mathbf s}(\mathbf{r}')\times\hat{\mathbf s}(\mathbf{r})$.
The third line is cross couplings between $x$ and $y$ components of the spins,
which will be referred to as the cross-spin interactions in what follows.
Finally, the last line serves as a local magnetic bias field, defining the quantization axis. 
  
The bias field and the coupling coefficients are position dependent and are related to the cavity mode function 
and the pump-field spatial profiles as 
$B_z(\mathbf{r})=\hbar\delta/2+\hbar|\Omega_1(\mathbf{r})|^2/2\Delta_1-\hbar|\Omega_2(\mathbf{r})|^2/2\Delta_2$,
$J_{\rm Heis}^{x/y}=\Re(c_1)\pm\Re(c_2)$, $J_{\rm DM}^{z}=-\Im(c_1)$, and $J_{\rm c}^{xy}=-\Im(c_2)$
with
\begin{align}
c_1(\mathbf{r}',\mathbf{r})=&
\left[\frac{1}{\Delta_1^2\tilde\Delta_c}\Omega_1(\mathbf{r}')\Omega_1^*(\mathbf{r})
      +\frac{1}{\Delta_2^2\tilde\Delta_c^*}\Omega_2^*(\mathbf{r}')\Omega_2(\mathbf{r})\right]\times\nonumber\\
      &2\hbar\mathscr{G}(x')\mathscr{G}(x),\nonumber\\
c_2(\mathbf{r}',\mathbf{r})=&\frac{1}{\Delta_1\Delta_2}
\left[\frac{1}{\tilde\Delta_c}\Omega_2(\mathbf{r}')\Omega_1^*(\mathbf{r})
      +\frac{1}{\tilde\Delta_c^*}\Omega_1^*(\mathbf{r}')\Omega_2(\mathbf{r})\right]\times\nonumber\\
      &2\hbar\mathscr{G}(x')\mathscr{G}(x).
\end{align}
Here $\Re$ ($\Im$) indicates the real (imaginary) part of a complex variable. 
We have assumed, without loss of generality, that $\mathscr{G}_0=\mathscr{G}_0^*$ is real
and have introduced 
$\tilde{\Delta}_c\equiv\Delta_c+i\kappa-\int \mathscr{G}^2(x)[\hat{n}_\downarrow(\mathbf{r})/\Delta_1
+\hat{n}_\uparrow(\mathbf{r})/\Delta_2]d\mathbf{r}$ for the shorthand.
The cavity-mediated interactions are infinite range as long as the laser
waists are much larger than the atomic cloud size as considered here.

Let us now consider the spatial profiles
$\Omega_{1,2}(y)=\Omega_{01,02}(e^{\pm ik_cy}+Re^{\mp ik_cy})/(1+R)$
for the lasers, where $0\leqslant R\leqslant1$ is the reflectivity of mirrors
retroreflecting the pump lasers. 
The amplitudes $\Omega_{01,02}$
are assumed to be real, with the balanced Raman condition 
$\eta_0\equiv\mathscr{G}_0\Omega_{01}/\Delta_1=\mathscr{G}_0\Omega_{02}/\Delta_2$ as in the experiment.
We work on the thermodynamic limit, where quantum fluctuations become negligible
and one can replace the photonic and atomic field operators with their 
corresponding quantum averages~\cite{piazza2013bose}:
$\hat{a}\to\langle\hat{a}\rangle\equiv\alpha$ and 
$\hat{\Psi}\to\langle\hat{\Psi}\rangle\equiv\Psi=(\psi_\uparrow,\psi_\downarrow)^\top$.
In this limit, we look for self-consistent solutions of the mean-field Hamiltonian corresponding to 
Eq.~\eqref{eq:single-particle-H-den}, endowed with steady-sate field amplitude $\alpha_{\rm ss}$ corresponding to 
Eq.~\eqref{eq:ss-a}~\cite{SM-spin-texture2018}. 

The mean-field phase diagram of the system in the $R$-$\eta_0$ plane is presented 
in Fig.~\ref{fig:domain_wall_ST}(a). Below the threshold 
pump-laser strength $\eta^{\rm c}_0(R)$, the steady state is a spin-polarized FM normal state 
where all the atoms are in the spin-up (as the bias field $B_z<0$ here), 
with no photon in the cavity. Above the threshold $\eta^{\rm c}_0(R)$,
the steady state of the system is a magnetically ordered chiral SS, domain-wall SS, or domain-wall AFM state depending
on $R$. In the following we examine these magnetic phases and their corresponding
spin Hamiltonian~\eqref{eq:eff-spin-H}.

\emph{Domain-wall AFM phase.}---The first
example we consider is the recent experiment of Ref.~\cite{Kroeze2018spin-texture},
where the pump lasers are both standing waves $\Omega_{1,2}(y)=\Omega_{01,02}\cos(k_cy)$ corresponding to $R=1$. 
In this case, the coefficients $c_1$ and $c_2$ are identical (i.e., $c_1=c_2$) and real.
Therefore, all the long-range spin-spin interactions vanish except
the $x$ component of the Heisenberg Hamiltonian, 
$\int J_{\rm Heis}^{x}(\mathbf{r}',\mathbf{r})\hat{s}_x(\mathbf{r}')\hat{s}_x(\mathbf{r}) d\mathbf{r}d\mathbf{r}'$
with the periodically modulated coupling strength 
$J_{\rm Heis}^{x}\propto\Re(\tilde\Delta_c)\cos(k_cx')\cos(k_cx)\cos(k_cy')\cos(k_cy)$~\cite{SM-spin-texture2018}.
Assuming $\Re(\tilde\Delta_c)<0$ and 
$\mathbf{r}-\mathbf{r}'=(m_x\hat{x}+m_y\hat{y})\lambda_c/2$
with $m_{x,y}$ being integers, $J_{\rm Heis}^{x}(\mathbf{r}',\mathbf{r})$ is positive (i.e., AFM)
when $m_x+m_y$ is odd, and it is negative (i.e., FM) when $m_x+m_y$ is even.

Above the laser-strength threshold $\eta_{\rm AFM}^{\rm c}\equiv\eta^{\rm c}_0(R=1)$, 
the system becomes unstable toward an ordered phase
with square density pattern, checkerboard domain-wall AFM spin texture, and finite cavity-photon number.
The $\mathbf{Z}_2$ symmetry of the system corresponding to the transformation 
$x\to x+\lambda_c/2$ and $\hat{a}\to-\hat{a}$ is spontaneously broken on the onset of the self-ordering phase transition
at $\eta_0=\eta_{\rm AFM}^{\rm c}$.
Different spin domains are separated by domain-wall lines, which are 1D topological defects.
A typical self-ordered domain-wall AFM spin texture 
is illustrated in Fig.~\ref{fig:domain_wall_ST}(b),
where the projection of the normalized local spin
$\tilde{\mathbf s}(\mathbf{r})\equiv\mathbf{s}(\mathbf{r})/s_n(\mathbf{r})$, with 
$\mathbf{s}(\mathbf{r})=\langle \hat{\mathbf s}(\mathbf{r})\rangle$ and 
$s_n(\mathbf{r})=\sqrt{s_x^2(\mathbf{r})+s_y^2(\mathbf{r})+s_z^2(\mathbf{r})}$,
in the $\tilde{s}_x$-$\tilde{s}_z$ plane is shown as a function of $\mathbf{r}$.
Note that the $y$ component of the spin is zero everywhere, $s_y(\mathbf{r})=0$~\cite{SM-spin-texture2018}. 
The $z$ component of the spin can also exhibit AFM behavior by properly introducing phases  
into the pump lasers (cf.\ $B_z$). In the 1D limit~\cite{Mivehvar2017disorder},
different domains are separated by domain-wall points, 0D topological defects~\cite{Braun2012-nanomagnetism}. 

\emph{Chiral SS phase.}---Let us now consider a
straightforward, though crucial, modification of the experiment of Ref.~\cite{Kroeze2018spin-texture}, 
where the driving lasers are assumed to be
counterpropagating running waves along the $y$ direction, $\Omega_{1,2}(y)=\Omega_{01,02}e^{\pm ik_cy}$,
corresponding to $R=0$. The coefficients $c_1$ and $c_2$ are 
now both complex and different from one another, $c_1\neq c_2$. Therefore, all the 
long-range spin-spin interactions are present in the effective Hamiltonian (\ref{eq:eff-spin-H}).
All the couplings have the same $\Re(\tilde{\Delta}_c)\cos(k_cx')\cos(k_cx)$
position dependence along the $x$ direction, but different modulations along the $y$ direction:
$J_{\rm Heis}^x\propto\cos(k_cy')\cos(k_cy)$, $J_{\rm Heis}^y\propto\sin(k_cy')\sin(k_cy)$,
$J_{\rm DM}^{z}\propto-\sin k_c(y'-y)$, and $J_{\rm c}^{xy}\propto\sin
k_c(y'+y)$~\cite{SM-spin-texture2018}.

Above the laser-strength threshold 
$\eta_{\rm SS}^{\rm c}\equiv\eta^{\rm c}_0(R=0)=
\pm\sqrt{-|\tilde\Delta_{0c}|^2(\omega_r+|\delta|/2)/N\Re(\tilde\Delta_{0c})}$ 
where $\tilde\Delta_{0c}=\langle \tilde\Delta_{c} \rangle$,
the spin-spin interactions
make the normal state unstable toward a magnetically self-ordered
state with finite photon number in the cavity~\cite{SM-spin-texture2018}. 
The DM and cross-spin interactions result in
an emergent transverse, conical SS state~\cite{Yoshida2012conical-spin-spiral,Haze2017chiral-spin}. 
The spirals solely appear in the $x$-$y$ plane
as the DM interaction has only the $z$ component and the cross-spin terms 
only couple the $x$ and $y$ components of the spins. 
The Heisenberg interactions favor magnetic domains as before. 
The steady state exhibits stripe-density patterns along the $x$ direction at minima of the 
cavity potential $x_{m_x}=m_x\lambda_c/2$. 

Figure~\ref{fig:chiral_ST}(a)-(c) illustrates a typical self-ordered chiral SS state.
The spin does a full $2\pi$ rotation in the
$\tilde{s}_x$-$\tilde{s}_y$ plane over one wave length $\lambda_c$ along the $y$ direction
due to the DM and cross-spin interactions.
This is clearly evident in Fig.~\ref{fig:chiral_ST}(c), which shows the projection of  
$\tilde{\mathbf s}(\mathbf{r})$ in the $\tilde{s}_x$-$\tilde{s}_y$ plane as a function of $\mathbf{r}$.
This can be understood by re-examining the sum of the DM and
cross-spin interactions, which is proportional to
\begin{align} \label{eq:DM-cs-H}
\Re(\tilde{\Delta}_c)&\int \cos(k_cx')\cos(k_cx)\Big[
\cos(k_cy')\sin(k_cy)\hat{s}_x(\mathbf{r}')\hat{s}_y(\mathbf{r})\nonumber\\
&+\sin(k_cy')\cos(k_cy)\hat{s}_y(\mathbf{r}')\hat{s}_x(\mathbf{r})\Big]d\mathbf{r}'d\mathbf{r}.
\end{align}
Along a stripe in the $y$ direction (i.e., $x'=x\neq m_x\lambda_c/4$), the coupling coefficients
change smoothly between negative and positive values; therefore, in
order to minimize these interactions the spin rotates fully in the $\tilde{s}_x$-$\tilde{s}_y$ plane along the $y$ axis
~\footnote{For instance, consider 
$\mathbf{r}'=(\lambda_c/2,0)$ and $\mathbf{r}=(\lambda_c/2,y)$,
where $y$ varies along the pump direction.
Then Eq.~\eqref{eq:DM-cs-H} simplifies to 
$\Re(\tilde{\Delta}_c)\int \sin(k_cy)\hat{s}_x(\lambda_c/2,0)\hat{s}_y(\lambda_c/2,y)dy$.
It is now easy to see from Fig.~\ref{fig:chiral_ST}(b) that the $x$ and $y$ components of the spin indeed change
along the $y$ axis in a way that minimizes this interaction (recall that $\Re(\tilde{\Delta}_c)<0$ here).}. 
For adjacent density stripes (i.e., $x'=x\pm\lambda_c/2$), the spin spirals are shifted 
by $\lambda_c/2$ along the $y$ axis due to the $x$-modulation of the couplings which introduces an extra minus 
sign~\footnote{For concreteness, consider $\mathbf{r}'=(\lambda_c/2,0)$ and $\mathbf{r}=(0,y)$,
where $y$ varies along the pump direction. 
Then Eq.~\eqref{eq:DM-cs-H} takes the form 
$-\Re(\tilde{\Delta}_c)\int \sin(k_cy)\hat{s}_x(\lambda_c/2,0)\hat{s}_y(0,y)dy$, favoring a spin
spiral at $x=0$ which is $\lambda_c/2$ shifted along the $y$ direction with respect to the spiral at $x'=\lambda_c/2$.}. 
These are also compatible with the Heisenberg interactions.
The discrete choice of the origin for the spin spirals is fixed at 
the onset of the self-ordering phase transition at $\eta_0=\eta_{\rm SS}^{\rm c}$ through the spontaneous 
$\mathbf{Z}_2$-symmetry breaking process. Here the continuous $U(1)$ screw symmetry of 
the system along the $y$ direction is broken explicitly by fixing the phases of the lasers.
Figure~\ref{fig:chiral_ST}(b) shows the projections of the normalized spin in the $\tilde{s}_x$-$\tilde{s}_z$
and $\tilde{s}_y$-$\tilde{s}_z$ planes, where the existence of magnetic domains due to the Heisenberg interactions are visible.
Note that the magnetic domains in the $\tilde{s}_x$-$\tilde{s}_z$
and $\tilde{s}_y$-$\tilde{s}_z$ planes are shifted by $\lambda_c/4$ along the $y$ axis with respect to each other,
consistent with the $y$ dependence of $J_{\rm Heis}^{x/y}$ given above~\cite{SM-spin-texture2018}.

\emph{Domain-wall SS phase.}---The coupling coefficient $J_{\rm Heis}^{y}$ approaches zero as $(1-R)^2$ 
and $\{J_{\rm DM}^{z}, J_{\rm c}^{xy} \}$ as $1-R^2$ when $R\to1$, while
the coupling coefficient $J_{\rm Heis}^{x}$ is independent of $R$~\cite{SM-spin-texture2018}.
Therefor, $J_{\rm Heis}^{x}$ is the dominant coupling coefficient for $0<R<1$, favoring magnetic domains. 
This is clearly obvious in Fig.~\ref{fig:chiral_ST}(d), which
displays the projection of the normalized spin texture in the $\tilde{s}_x$-$\tilde{s}_y$ plane for $R=0.2$.
Although the spin still does a full $2\pi$ rotation in the $\tilde{s}_x$-$\tilde{s}_y$ plane over one wave 
length $\lambda_c$ along the $y$ direction, the $y$ component of the spin is considerably suppressed 
compared to the $R=0$ case shown in Fig.~\ref{fig:chiral_ST}(c). Therefore, we refer to this phase as the
``domain-wall SS" phase.

\emph{Concluding remarks.}---We have shown that a variety of long-range spin
models can be implemented using driven atoms coupled to an optical
cavity, by properly choosing the driving electromagnetic modes.
As illustrations, 
we demonstrated the emergence of the experimentally relevant domain-wall AFM/SS and chiral SS orders.
The transition between these two magnetic phases can be explored in a single
state-of-the-art experimental setup~\cite{Landini2018spin-texture, Kroeze2018spin-texture},
by continuously tuning the reflectively $R$ of retroreflectors. The physics presented here
remains qualitatively the same for a wide range of parameters, including large cavity bandwidth $\kappa\gg\omega_r$ limit. 

As possible extensions of our scheme, we mention that
general laser configurations $\Omega_{1,2}(\mathbf{r})$,
where $\mathbf{r}$ is not restricted only to the $y$ axis, 
would allow to implement more complex spatial modulations of
the spin-spin interactions. This feature, together with the
tuneability of the interaction range in multimode cavities~\cite{kroeze_tuneable_2018}
and the inclusion of short-range spin-spin interactions due to contact collisional interactions, 
should introduce frustration~\cite{Balents2010spin-liquid}. 
Another generalization is to apply an extra pump laser with the same polarization as the cavity
mode, in order to further induce the transitions $\downarrow\>\leftrightarrow 1$ and $\uparrow\>\leftrightarrow 2$. 
This would result in long-range spin-spin interactions of the form
$s_z(\mathbf{r}')s_z(\mathbf{r})$, 
$s_x(\mathbf{r}')s_z(\mathbf{r})$, and $s_y(\mathbf{r}')s_z(\mathbf{r})$, 
yielding additional components of the Heisenberg and DM interactions. 
This should allow the realization of topological chiral phases such as skyrmions.
It is also worth mentioning that the linear cavity  can be replaced by a ring cavity with running-wave
modes, providing another complex degree of freedom to tune the
cavity-mediated spin-spin interactions with a continuous $U(1)$ spatial
symmetry~\cite{Mivehvar2015stripe-phase, Mivehvar2018supersolid, Ostermsnn2018spin-spiral}.
As the proposed setups can be implemented in several laboratories 
with straightforward modifications to the state of the art, 
the framework of cavity QED appears to be a very promising candidate 
for the controlled simulation of quantum magnetism.

\begin{acknowledgments}
We are grateful to N. Dogra, T. Donner, M. Landini, and R. M. Kroeze for fruitful discussions.
FM is supported by the Lise-Meitner Fellowship M2438-NBL 
and the international FWF-ANR grand, No.\ I3964-N27.
HR acknowledges support from the Austrian Science Fund FWF 
through No.\ I1697-N27.
\end{acknowledgments}

\bibliography{2BEC_lin_cav_2D_spin_texture}

\newpage
\widetext
\setcounter{equation}{0}
\setcounter{figure}{0}
\renewcommand{\theequation}{S\arabic{equation}}
\renewcommand{\thefigure}{S\arabic{figure}}

\section{Supplemental Material}

Here we present the details of the adiabatic elimination of the atomic excited states and the cavity field, 
the derivation of the effective spin Hamiltonian, the details of linear stability analysis, some additional
spin textures, and the details of our numerical approach.

\section{Adiabatic Elimination of the Atomic Excited States}
 
The system presented in the main text is described, in the dipole and rotating wave approximations, 
by the single-particle Hamiltonian density:
\begin{align} \label{eq:1-particle-H-density}
\hat{\mathcal H}_4=\frac{\mathbf{p}^2}{2m}I_{4\times4}
+\sum_{\tau=\{\uparrow,1,2\}}\hbar\omega_\tau\sigma_{\tau \tau}
+\hbar\omega_c\hat{a}^\dagger\hat{a}
+\hbar\Big[\mathscr{G}(x)\hat{a}(\sigma_{1\downarrow}+\sigma_{2\uparrow}) 
+\Omega_1(\mathbf{r})e^{-i\omega_{p1}t}\sigma_{1\uparrow}
+\Omega_2(\mathbf{r})e^{-i\omega_{p2}t}\sigma_{2\downarrow}
+ \text{H.c.}\Big],
\end{align}
where $\sigma_{\tau \tau'}=\ket{\tau}\bra{\tau'}$, H.c.\ stands for the Hermitian conjugate, and $I_{4 \times 4}$ is the identity matrix in the internal atomic-state space. The single-particle Hamiltonian density~\eqref{eq:1-particle-H-density} can be brought into a time-independent form 
\begin{align} \label{eq:1-particle-H-density-rot-frame}
\hat{\tilde{\mathcal H}}_4&=\mathscr{U}\hat{\mathcal H}_4\mathscr{U}^\dag+i\hbar(\partial_t \mathscr{U})\mathscr{U}^\dag\nonumber\\
&=\frac{\mathbf{p}^2}{2M}I_{4\times4}
+\hbar\delta\sigma_{\uparrow\uparrow}
-\hbar\Delta_1\sigma_{11}-\hbar\Delta_2\sigma_{22}
-\hbar\Delta_c\hat{a}^\dagger\hat{a}
+\hbar\Big[\mathscr{G}(x)\hat{a}(\sigma_{1\downarrow}+\sigma_{2\uparrow}) 
+\Omega_1(\mathbf{r})\sigma_{1\uparrow}+\Omega_2(\mathbf{r})\sigma_{2\downarrow}
+ \text{H.c.}\Big],
\end{align}
using the unitary operator
\begin{align} \label{U-T}
\mathscr{U}=\exp{\left\{i\left[\left(\frac{\omega_{p2}-\omega_{p1}}{2}\right)\sigma_{\uparrow\uparrow}
+\left(\frac{\omega_{p1}+\omega_{p2}}{2}\right)\left(\sigma_{11}+\hat{a}^\dagger\hat{a}\right)
+\omega_{p2}\sigma_{22}\right]t\right\}}.
\end{align}
Here, we have defined 
$\Delta_1\equiv(\omega_{p1}+\omega_{p2})/2-\omega_1$,
$\Delta_2\equiv\omega_{p2}-\omega_2$,
and $\Delta_c\equiv(\omega_{p1}+\omega_{p2})/2-\omega_c$ as the atomic and cavity detunings 
with respect to the pump lasers, respectively, and $\delta\equiv\omega_\uparrow-(\omega_{p2}-\omega_{p1})/2+{B}_{\rm ext}$
is the relative two-photon detuning. Here we have also included an external magnetic field ${B}_{\rm ext}$ to tune the internal
atomic levels.

The corresponding many-body Hamiltonian takes the form,
\begin{align} \label{eq:many-body-H}
\hat{\tilde H}_4=\int \hat\Psi^\dag_4(\mathbf{r})\hat{\tilde{\mathcal H}}_4
\hat\Psi_4(\mathbf{r})d\mathbf{r}+\hat{H}_{\text{int}}, 
\end{align}
where $\hat\Psi_4(\mathbf{r})=(\hat\psi_\downarrow,\hat\psi_\uparrow,\hat\psi_1,\hat\psi_2)^{\mathsf{T}}$ 
are the bosonic atomic field operators 
satisfying the usual bosonic commutation relation 
$[\hat\psi_\tau(\mathbf{r}),\hat\psi_{\tau'}^\dag(\mathbf{r}')]=\delta_{\tau,\tau'}\delta(\mathbf{r}-\mathbf{r}')$. 
The interaction Hamiltonian $\hat{H}_{\text{int}}$ accounts for two-body contact interactions between atoms
and will be omitted in the following.

The Heisenberg equations of motion of the photonic and atomic field operators can be obtained using 
the many-body Hamiltonian~\eqref{eq:many-body-H},
\begin{align} \label{eq:Heisenberg-eqs-motion}
i\hbar\frac{\partial}{\partial t}\hat{a}&=-\hbar\Delta_c\hat{a}-i\hbar\kappa\hat{a}
+\hbar\int \mathscr{G}^*(x)\left[\hat\psi_\downarrow^\dag(\mathbf{r})\hat\psi_{1}(\mathbf{r})
+\hat\psi_\uparrow^\dag(\mathbf{r})\hat\psi_{2}(\mathbf{r})\right] d\mathbf{r},\nonumber\\
i\hbar\frac{\partial}{\partial t}\hat\psi_\downarrow&=-\frac{\hbar^2}{2M}\boldsymbol\nabla^2 \hat\psi_\downarrow
+\hbar\mathscr{G}^*(x)\hat{a}^\dag\hat\psi_1+\hbar\Omega_2^*(\mathbf{r})\hat\psi_2,\nonumber\\
i\hbar\frac{\partial}{\partial t}\hat\psi_\uparrow&=\left(-\frac{\hbar^2}{2M}\boldsymbol\nabla^2+\hbar\delta\right)\hat\psi_\uparrow
+\hbar\mathscr{G}^*(x)\hat{a}^\dag\hat\psi_2+\hbar\Omega_1^*(\mathbf{r})\hat\psi_1,\nonumber\\
i\hbar\frac{\partial}{\partial t}\hat\psi_1&=\left(-\frac{\hbar^2}{2M}\boldsymbol\nabla^2-\hbar\Delta_1\right)\hat\psi_1
+\hbar\mathscr{G}(x)\hat{a}\hat\psi_\downarrow+\hbar\Omega_1(\mathbf{r})\hat\psi_\uparrow,\nonumber\\
i\hbar\frac{\partial}{\partial t}\hat\psi_2&=\left(-\frac{\hbar^2}{2M}\boldsymbol\nabla^2-\hbar\Delta_2\right)\hat\psi_2
+\hbar\mathscr{G}(x)\hat{a}\hat\psi_\uparrow+\hbar\Omega_2(\mathbf{r})\hat\psi_\downarrow,
\end{align}
where we have phenomenologically included the decay of the cavity mode $-i\hbar\kappa\hat{a}$. 
Let us now assume that atomic detunings $\Delta_{j}$ are large so that the atomic field operators
$\{\hat\psi_1,\hat\psi_2\}$ of the excited states reach steady states $\partial_t \hat\psi_{1,\rm ss}=\partial_t \hat\psi_{2,\rm ss}=0$ 
very fast and we can therefore adiabatically 
eliminate their dynamics. Omitting the kinetic energies in comparison to $-\hbar\Delta_1$ and $-\hbar\Delta_2$, 
we obtain the steady-state atomic field operators of the excited states,
\begin{align} \label{eq:ss-atomic-excited-field-op}
\hat\psi_{1,\text{ss}}=\frac{1}{\Delta_1}
\left[\mathscr{G}(x)\hat{a}\hat\psi_\downarrow+\Omega_1(\mathbf{r})\hat\psi_\uparrow\right],\nonumber\\
\hat\psi_{2,\text{ss}}=\frac{1}{\Delta_2}
\left[\mathscr{G}(x)\hat{a}\hat\psi_\uparrow+\Omega_2(\mathbf{r})\hat\psi_\downarrow\right].
\end{align}  
Substituting the steady-state atomic field operators of the excited states~\eqref{eq:ss-atomic-excited-field-op} 
in the rest of the Heisenberg equations of motion~\eqref{eq:Heisenberg-eqs-motion} yields a set of effective equations for the photonic and ground-state atomic field operators,
\begin{align}  \label{eqSM:eff-Heisenberg-eqs-motion}
i\hbar\frac{\partial}{\partial t}\hat{a}&=\hbar\left\{-\Delta_c-i\kappa
+\int |\mathscr{G}(x)|^2
\left[\frac{1}{\Delta_1}\hat{n}_\downarrow(\mathbf{r})
+\frac{1}{\Delta_2}\hat{n}_\uparrow(\mathbf{r})\right] d\mathbf{r}\right\}\hat{a}
+\hbar\int \mathscr{G}^*(x)\left[\frac{1}{\Delta_1}\Omega_1(\mathbf{r})\hat{s}_-(\mathbf{r})
+\frac{1}{\Delta_2}\Omega_2(\mathbf{r})\hat{s}_+(\mathbf{r})\right]d\mathbf{r},\nonumber\\
i\hbar\frac{\partial}{\partial t}\hat\psi_\uparrow&=\left[-\frac{\hbar^2}{2M}\boldsymbol\nabla^2
+\hbar\delta+\hat{V}_\uparrow(\mathbf{r})\right]\hat\psi_\uparrow
+\hbar\hat{\Omega}_{\rm R}(\mathbf{r})\hat\psi_\downarrow,\nonumber\\
i\hbar\frac{\partial}{\partial t}\hat\psi_\downarrow&=\left[-\frac{\hbar^2}{2M}\boldsymbol\nabla^2
+\hat{V}_\downarrow(\mathbf{r})\right]\hat\psi_\downarrow
+\hbar\hat{\Omega}_{\rm R}^\dag(\mathbf{r})\hat\psi_\uparrow,
\end{align}
with the potential operators 
$\hat{V}_\uparrow(\mathbf{r})=\hbar\hat{a}^\dag\hat{a}|\mathscr{G}(x)|^2/\Delta_2+\hbar|\Omega_1(\mathbf{r})|^2/\Delta_1$ and
$\hat{V}_\downarrow(\mathbf{r})=\hbar\hat{a}^\dag\hat{a}|\mathscr{G}(x)|^2/\Delta_1+\hbar|\Omega_2(\mathbf{r})|^2/\Delta_2$,
and the two-photon Raman-Rabi coupling operator 
$\hat{\Omega}_{\rm R}(\mathbf{r})=\Omega_1^*(\mathbf{r})\hat{a}\mathscr{G}(x)/\Delta_1
+\hat{a}^\dag\mathscr{G}^*(x)\Omega_2(\mathbf{r})/\Delta_2$.
The many-body Hamiltonian 
$\hat{H}=-\hbar\Delta_c\hat{a}^\dag\hat{a}+\int \hat\Psi^\dag(\mathbf{r})\hat{\mathcal H}_0\hat\Psi(\mathbf{r})d\mathbf{r}$
can be easily read out, with $\hat\Psi(\mathbf{r})=(\hat\psi_\uparrow,\hat\psi_\downarrow)^{\mathsf{T}}$ and 
the effective single-particle Hamiltonian density $\hat{\mathcal H}_0$ given in Eq.~(1) in the main text.

\section{Adiabatic Elimination of the Cavity Mode: The Effective Spin Hamiltonian}

The steady-state cavity field operator $i\hbar\partial_t \hat{a}_{\rm ss}=[\hat{a}_{\rm ss},\hat{H}]=0$,
\begin{align} \label{eqSM:ss-a}
\hat{a}_{\rm ss}=\frac{1}{\tilde\Delta_c}
\int \mathscr{G}^*(x)\left[\frac{1}{\Delta_1}\Omega_1(\mathbf{r})\hat{s}_-(\mathbf{r})
+\frac{1}{\Delta_2}\Omega_2(\mathbf{r})\hat{s}_+(\mathbf{r})\right]d\mathbf{r},
\end{align}
with $\tilde{\Delta}_c\equiv
\Delta_c+i\kappa-\int |\mathscr{G}(x)|^2[\hat{n}_\downarrow(\mathbf{r})/\Delta_1
+\hat{n}_\uparrow(\mathbf{r})/\Delta_2]d\mathbf{r}$ for the shorthand,
can be substituted in the many-body Hamiltonian $\hat{H}$
to yield an effective atomic Hamiltonian,
\begin{align}
\hat{H}_{\rm eff}=\int \hat\Psi^\dag(\mathbf{r})
 \left[\frac{\mathbf{p}^2}{2M}
 +\frac{\hbar\delta}{2}+\frac{\hbar}{2\Delta_1}|\Omega_1(\mathbf{r})|^2
 +\frac{\hbar}{2\Delta_2}|\Omega_2(\mathbf{r})|^2 \right]
 I_{2\times2}
\hat\Psi(\mathbf{r})d\mathbf{r}
+\hat{H}_{\rm spin}.
\end{align}
The Hamiltonian $\hat{H}_{\rm spin}$ is an effective long-range spin Hamiltonian stemmed from 
photon-mediated interactions and it reads,
\begin{align} \label{eqSM:spin-H}
\hat{H}_{\rm spin}&=
2\hbar\int \Bigg\{
\Re\left[\frac{1}{\Delta_1^2\tilde\Delta_c}\mathscr{G}^*(x')\mathscr{G}(x)\Omega_1(\mathbf{r}')\Omega_1^*(\mathbf{r})
          +\frac{1}{\Delta_2^2\tilde\Delta_c^*}\mathscr{G}(x')\mathscr{G}^*(x)\Omega_2^*(\mathbf{r}')\Omega_2(\mathbf{r})\right]
          \left[\hat{s}_x(\mathbf{r}')\hat{s}_x(\mathbf{r})+\hat{s}_y(\mathbf{r}')\hat{s}_y(\mathbf{r})\right] \nonumber\\
&+\frac{1}{\Delta_1\Delta_2}
\Re\left[\frac{1}{\tilde\Delta_c}\mathscr{G}^*(x')\mathscr{G}(x)\Omega_2(\mathbf{r}')\Omega_1^*(\mathbf{r})
          +\frac{1}{\tilde\Delta_c^*}\mathscr{G}(x')\mathscr{G}^*(x)\Omega_1^*(\mathbf{r}')\Omega_2(\mathbf{r})\right]
          \left[\hat{s}_x(\mathbf{r}')\hat{s}_x(\mathbf{r})-\hat{s}_y(\mathbf{r}')\hat{s}_y(\mathbf{r})\right]\nonumber\\
&-\Im\left[\frac{1}{\Delta_1^2\tilde\Delta_c}\mathscr{G}^*(x')\mathscr{G}(x)\Omega_1(\mathbf{r}')\Omega_1^*(\mathbf{r})
          +\frac{1}{\Delta_2^2\tilde\Delta_c^*}\mathscr{G}(x')\mathscr{G}^*(x)\Omega_2^*(\mathbf{r}')\Omega_2(\mathbf{r})\right]
          \left[\hat{s}_x(\mathbf{r}')\hat{s}_y(\mathbf{r})-\hat{s}_y(\mathbf{r}')\hat{s}_x(\mathbf{r})\right] \nonumber\\
&-\frac{1}{\Delta_1\Delta_2}
\Im\left[\frac{1}{\tilde\Delta_c}\mathscr{G}^*(x')\mathscr{G}(x)\Omega_2(\mathbf{r}')\Omega_1^*(\mathbf{r})
          +\frac{1}{\tilde\Delta_c^*}\mathscr{G}(x')\mathscr{G}^*(x)\Omega_1^*(\mathbf{r}')\Omega_2(\mathbf{r})\right]
          \left[\hat{s}_x(\mathbf{r}')\hat{s}_y(\mathbf{r})+\hat{s}_y(\mathbf{r}')\hat{s}_x(\mathbf{r})\right]                           
          \Bigg\}d\mathbf{r}d\mathbf{r}'\nonumber\\
&+\int \left[\frac{\hbar\delta}{2}+\frac{\hbar}{2\Delta_1}|\Omega_1(\mathbf{r})|^2-\frac{\hbar}{2\Delta_2}|\Omega_2(\mathbf{r})|^2 \right]
                \hat{s}_z(\mathbf{r}) d\mathbf{r}
+\sum_{j=0}^3\mathcal{O}(\frac{1}{\Delta_1^{3-j}\Delta_2^j}).
\end{align}
The terms arising from the free cavity Hamiltonian $-\hbar\Delta_c\hat{a}^\dag\hat{a}$ and the cavity potentials $\hbar\hat{a}^\dag\hat{a}|\mathscr{G}(x)|^2/\Delta_{1,2}$, as well as from the non-commutation of the steady-state cavity field operator~\eqref{eqSM:ss-a} with the atomic field operators, have been discarded. The spin Hamiltonian~\eqref{eqSM:spin-H} can be recast in a more compact way,
\begin{align}\label{eqSM:eff-spin-H}
\hat{H}_{\rm spin}&=\int \Big\{
J_{\rm Heis}^{x}(\mathbf{r}',\mathbf{r})\hat{s}_x(\mathbf{r}')\hat{s}_x(\mathbf{r})
+J_{\rm Heis}^{y}(\mathbf{r}',\mathbf{r})\hat{s}_y(\mathbf{r}')\hat{s}_y(\mathbf{r})
+J_{\rm DM}^{z}(\mathbf{r}',\mathbf{r})\left[\hat{s}_x(\mathbf{r}')\hat{s}_y(\mathbf{r})-\hat{s}_y(\mathbf{r}')\hat{s}_x(\mathbf{r})\right]\nonumber\\
&+J_{\rm c}^{xy}(\mathbf{r}',\mathbf{r})\left[\hat{s}_x(\mathbf{r}')\hat{s}_y(\mathbf{r})+\hat{s}_y(\mathbf{r}')\hat{s}_x(\mathbf{r})\right]
\Big\}d\mathbf{r}d\mathbf{r}'
+\int B_z(\mathbf{r})\hat{s}_z(\mathbf{r}) d\mathbf{r},
\end{align}
where
$B_z=\hbar\delta/2+\hbar|\Omega_1(\mathbf{r})|^2/2\Delta_1-\hbar|\Omega_2(\mathbf{r})|^2/2\Delta_2$,
$J_{\rm Heis}^{x/y}=\Re(c_1)\pm\Re(c_2)$, $J_{\rm DM}^{z}=-\Im(c_1)$, and $J_{\rm c}^{xy}=-\Im(c_2)$ 
with
\begin{align}
c_1(\mathbf{r}',\mathbf{r})&=
2\hbar\mathscr{G}(x')\mathscr{G}(x)
\left[\frac{1}{\Delta_1^2\tilde\Delta_c}\Omega_1(\mathbf{r}')\Omega_1^*(\mathbf{r})
      +\frac{1}{\Delta_2^2\tilde\Delta_c^*}\Omega_2^*(\mathbf{r}')\Omega_2(\mathbf{r})\right],\nonumber\\
c_2(\mathbf{r}',\mathbf{r})&=\frac{2\hbar}{\Delta_1\Delta_2}\mathscr{G}(x')\mathscr{G}(x)
\left[\frac{1}{\tilde\Delta_c}\Omega_2(\mathbf{r}')\Omega_1^*(\mathbf{r})
      +\frac{1}{\tilde\Delta_c^*}\Omega_1^*(\mathbf{r}')\Omega_2(\mathbf{r})\right].
\end{align}
Here we have assumed, without loss of generality, that $\mathscr{G}_0=\mathscr{G}_0^*$ is real.

\subsection{The Coupling Coefficients}

\begin{figure}[t!]
\centering
\includegraphics [width=0.75\textwidth]{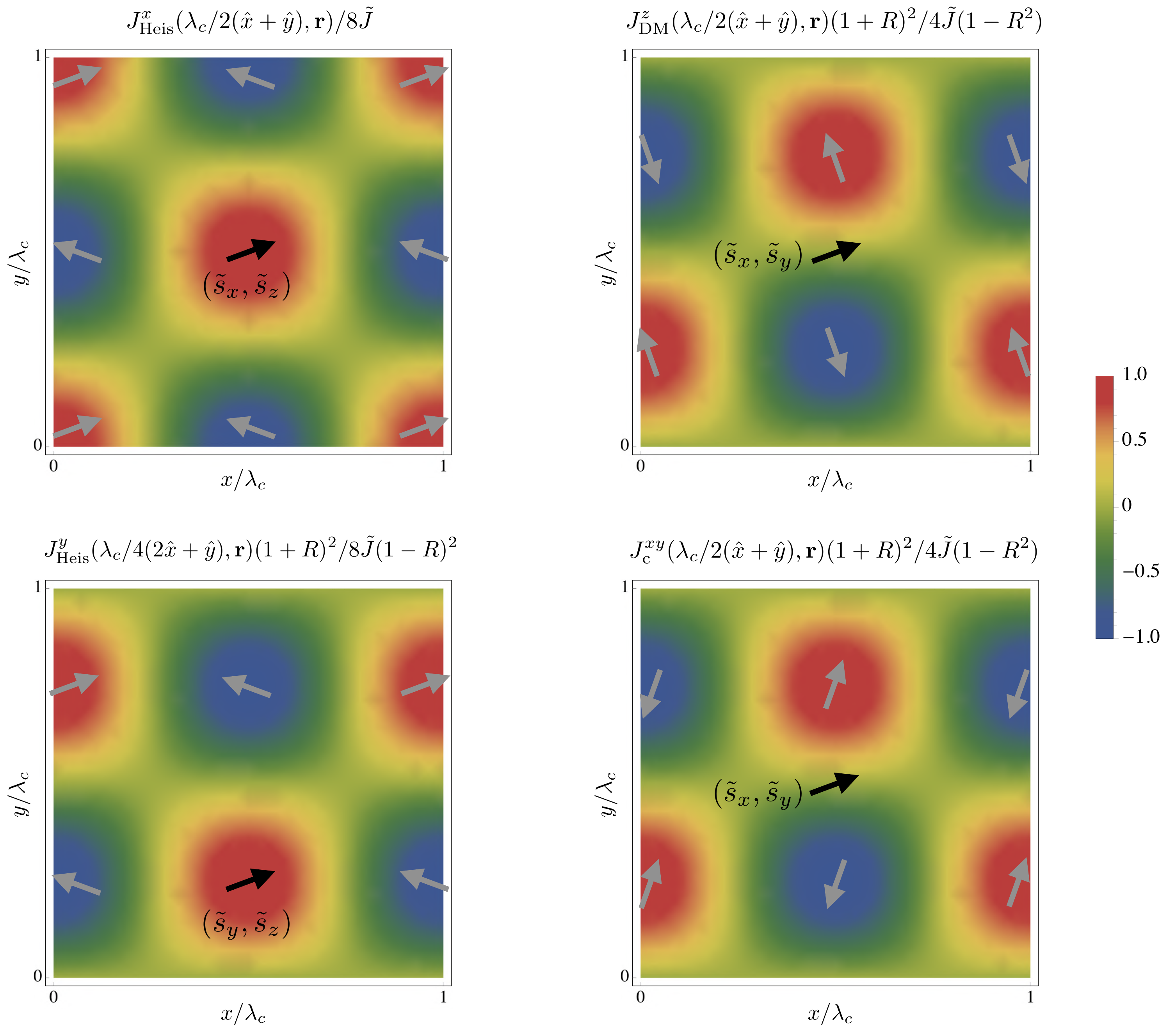}
\caption{The spatial dependence of the cavity-induced spin-spin interaction coefficients. 
The graphs are the rescaled unitless (i.e., the spatial part of) 
$J_{\rm Heis}^{x}(\lambda_c/2\hat{x}+\lambda_c/2\hat{y},\mathbf{r})$,
$J_{\rm Heis}^{y}(\lambda_c/2\hat{x}+\lambda_c/4\hat{y},\mathbf{r})$,
$J_{\rm DM}^{z}(\lambda_c/2\hat{x}+\lambda_c/2\hat{y},\mathbf{r})$,
and $J_{\rm c}^{xy}(\lambda_c/2\hat{x}+\lambda_c/2\hat{y},\mathbf{r})$
as a function of $\mathbf{r}$ for fixed $\mathbf{r}'$. 
Here $\tilde{J}\equiv\hbar\Re(\tilde\Delta_c)\eta_0^2/|\tilde\Delta_c|^2$ as a shorthand.
In each figure, the black arrow indicates an arbitrary spin at position
$\mathbf{r}'$, and the gray arrows show the
spin orientations which minimize the corresponding spin-spin interaction. 
Note that $\tilde{J}<0$ for
our chosen parameters as in Figs.~2 and 3 in the main text.} 
\label{figSM:Js}
\end{figure}

We now consider the spatial profiles $\Omega_{1,2}(y)=\Omega_{01,02}(e^{\pm ik_cy}+Re^{\mp ik_cy})/(1+R)$ for the pump lasers, with $0\leqslant R\leqslant1$ being the retroreflectivity of mirrors retroreflecting the pump lasers. The coefficients $c_{1,2}$ take the form,
\begin{align} \label{eqSM:cs-gerneral}
c_1(\mathbf{r}',\mathbf{r})&=\frac{4\hbar\Re(\tilde\Delta_c)\eta_0^2}{|\tilde\Delta_c|^2(1+R)^2}
\cos(k_cx')\cos(k_cx)
\left[e^{ik_c(y'-y)}+R^2e^{-ik_c(y'-y)}+2R\cos k_c(y'+y) \right],\nonumber\\
c_2(\mathbf{r}',\mathbf{r})&=\frac{4\hbar\Re(\tilde\Delta_c)\eta_0^2}{|\tilde\Delta_c|^2(1+R)^2}
\cos(k_cx')\cos(k_cx)
\left[e^{-ik_c(y'+y)}+R^2e^{ik_c(y'+y)}+2R\cos k_c(y'-y) \right],
\end{align}
where the amplitudes $\Omega_{01}$ and $\Omega_{02}$ have been assumed to be real 
with the balanced Raman condition $\eta_0\equiv\mathscr{G}_0\Omega_{01}/\Delta_1=\mathscr{G}_0\Omega_{02}/\Delta_2$. 
Using these coefficients~\eqref{eqSM:cs-gerneral}, the spin-spin coupling coefficients can be readily obtained as
\begin{align} \label{eqSM:Js-gerneral}
J_{\rm Heis}^x(\mathbf{r}',\mathbf{r})&=\frac{8\hbar\Re(\tilde\Delta_c)\eta_0^2}{|\tilde\Delta_c|^2}
\cos(k_cx')\cos(k_cx)\cos(k_cy')\cos(k_cy),\nonumber\\
J_{\rm Heis}^y(\mathbf{r}',\mathbf{r})&=\frac{8\hbar\Re(\tilde\Delta_c)\eta_0^2}{|\tilde\Delta_c|^2}
\frac{(1-R)^2}{(1+R)^2}
\cos(k_cx')\cos(k_cx)\sin(k_cy')\sin(k_cy),\nonumber\\
J_{\rm DM}^{z}(\mathbf{r}',\mathbf{r})&=-\frac{4\hbar\Re(\tilde\Delta_c)\eta_0^2}{|\tilde\Delta_c|^2}
\frac{1-R^2}{(1+R)^2}
\cos(k_cx')\cos(k_cx)\sin k_c(y'-y),\nonumber\\
J_{\rm c}^{xy}(\mathbf{r}',\mathbf{r})&=\frac{4\hbar\Re(\tilde\Delta_c)\eta_0^2}{|\tilde\Delta_c|^2}
\frac{1-R^2}{(1+R)^2}
\cos(k_cx')\cos(k_cx)\sin k_c(y'+y).
\end{align}
The coupling coefficient $J_{\rm Heis}^x$ is independent of the reflectivity $R$, while all the other coupling coefficients depend on $R$. 
For $R=1$, corresponding to standing-wave pump lasers $\Omega_{1,2}(y)=\Omega_{01,02}\cos(k_cy)$ along the $y$ direction, all the coupling coefficients are zero, saving for $J_{\rm Heis}^x$. While for $0\leqslant R<1$, all the coupling coefficients are present and compete with each other. The spatial dependence of the coupling coefficients~\eqref{eqSM:Js-gerneral} are plotted in Fig.~\ref{figSM:Js} as a function of $\mathbf{r}$ for fixed $\mathbf{r}'$. Note that both the DM and cross-spin interactions favor chiral spin-spiral textures. 
 
\section{Linearized Equations}

In this section we present the stability analysis of the mean-field solutions.  
To this end, we linearize the Heisenberg equations of motion~\eqref{eqSM:eff-Heisenberg-eqs-motion} of the field operators
$\hat{\psi}_\tau(\mathbf{r},t)=e^{-i\mu t/\hbar}[\psi_\tau(\mathbf{r})+\delta\hat{\psi}_\tau(\mathbf{r},t)]$ and
$\hat{a}(t)=\alpha+\delta\hat{a}(t)$ around the mean-field solutions $\psi_\tau(\mathbf{r})$ and $\alpha$:
\begin{align} \label{eqSM:linearized_eq}
i\frac{\partial}{\partial t} \delta\hat{a}&=-\tilde{\Delta}_{0c}\delta\hat{a}
+\int \left(\mathcal{A}_\uparrow\delta\hat{\psi}_\uparrow+\mathcal{A}_{\uparrow*}\delta\hat{\psi}_\uparrow^*
+\mathcal{A}_\downarrow\delta\hat{\psi}_\downarrow+\mathcal{A}_{\downarrow*}\delta\hat{\psi}_\downarrow^*\right)d\mathbf{r},\nonumber\\
i\frac{\partial}{\partial t}\delta\hat{\psi}_\uparrow&=\frac{1}{\hbar}\left[-\frac{\hbar^2}{2M}\boldsymbol\nabla^2
+\hbar\delta+V_\uparrow-\mu\right]\delta\hat\psi_\uparrow
+\Omega_{\rm R}\delta\hat\psi_\downarrow
+\mathcal{A}_{\uparrow}^*\delta\hat{a}+\mathcal{A}_{\uparrow*}\delta\hat{a}^*,\nonumber\\
i\frac{\partial}{\partial t}\delta\hat{\psi}_\downarrow&=\frac{1}{\hbar}\left[-\frac{\hbar^2}{2M}\boldsymbol\nabla^2
+V_\downarrow-\mu\right]\delta\hat\psi_\downarrow
+\Omega_{\rm R}^*\delta\hat\psi_\uparrow
+\mathcal{A}_{\downarrow}^*\delta\hat{a}+\mathcal{A}_{\downarrow*}\delta\hat{a}^*,
\end{align}
where $\{\delta\hat{\psi}_\tau,\delta\hat{a}\}$ are quantum fluctuations of the atomic and photonic fields.
Here, $\mu$ is the chemical potential, 
$\tilde{\Delta}_{0c}=\langle\tilde{\Delta}_{c} \rangle=
\Delta_c+i\kappa-\int \mathscr{G}^2(x)[n_\downarrow(\mathbf{r})/\Delta_1
+n_\uparrow(\mathbf{r})/\Delta_2]d\mathbf{r}$,
$V_\tau(\mathbf{r})=\langle \hat{V}_\tau(\mathbf{r}) \rangle$,
$\Omega_{\rm R}(\mathbf{r})=\langle \hat{\Omega}_{\rm R}(\mathbf{r}) \rangle$, 
and we have introduced the following notations for the shorthand:
 \begin{align} \label{eqSM:A-coeff}
\mathcal{A}_\uparrow(\mathbf{r})&=\frac{|\mathscr{G}|^2}{\Delta_2}\alpha\psi_\uparrow^*
+\frac{\mathscr{G}^*\Omega_1}{\Delta_1}\psi_\downarrow^*,\nonumber\\
\mathcal{A}_{\uparrow*}(\mathbf{r})&=\frac{|\mathscr{G}|^2}{\Delta_2}\alpha\psi_\uparrow
+\frac{\mathscr{G}^*\Omega_2}{\Delta_2}\psi_\downarrow,\nonumber\\
\mathcal{A}_\downarrow(\mathbf{r})&=\frac{|\mathscr{G}|^2}{\Delta_1}\alpha\psi_\downarrow^*
+\frac{\mathscr{G}^*\Omega_2}{\Delta_2}\psi_\uparrow^*,\nonumber\\
\mathcal{A}_{\downarrow*}(\mathbf{r})&=\frac{|\mathscr{G}|^2}{\Delta_1}\alpha\psi_\downarrow
+\frac{\mathscr{G}^*\Omega_1}{\Delta_1}\psi_\uparrow.
 \end{align}
Assuming $\delta\hat{\psi}_\tau(\mathbf{r},t)=\delta\hat{\psi}_{\tau+}(\mathbf{r})e^{-i\omega t}
+\delta\hat{\psi}_{\tau-}^*(\mathbf{r})e^{i\omega^* t}$ and 
$\delta\hat{a}(t)=\delta\hat{a}_+e^{-i\omega t}+\delta\hat{a}_-^*e^{i\omega^* t}$ for the quantum fluctuations,
the linearized equations~\eqref{eqSM:linearized_eq} can be recast in a matrix form
\begin{align} \label{eqSM:Bog-eq}
\omega\mathbf{f}=\mathbf{M}_{\rm B}\mathbf{f},
\end{align}
where 
$\mathbf{f}=(
\delta\hat{a}_+,\delta\hat{a}_-,
\delta\hat{\psi}_{\uparrow+},\delta\hat{\psi}_{\uparrow-},
\delta\hat{\psi}_{\downarrow+},\delta\hat{\psi}_{\downarrow-})^\mathsf{T}$
and 
\begin{align} \label{eqSM:Bog-matrix}
\renewcommand*{\arraystretch}{1.25}
\mathbf{M}_{\rm B}=
\begin{pmatrix}
-\tilde\Delta_{0c} & 0 & \int d\mathbf{r}\mathcal{A}_\uparrow &  \int d\mathbf{r}\mathcal{A}_{\uparrow*} 
& \int d\mathbf{r}\mathcal{A}_\downarrow & \int d\mathbf{r}\mathcal{A}_{\downarrow*} \\
0 & \tilde\Delta_{0c}^* & -\int d\mathbf{r}\mathcal{A}_{\uparrow*}^*  & -\int d\mathbf{r}\mathcal{A}_\uparrow^* 
& -\int d\mathbf{r}\mathcal{A}_{\downarrow*}^* & -\int d\mathbf{r}\mathcal{A}_\downarrow^*  \\
\mathcal{A}_\uparrow^* & \mathcal{A}_{\uparrow*}  & (\mathcal{H}_{0\uparrow}-\mu)/\hbar & 0 & \Omega_{\rm R} & 0 \\
-\mathcal{A}_{\uparrow*}^* & -\mathcal{A}_\uparrow & 0 & -(\mathcal{H}_{0\uparrow}-\mu)/\hbar & 0 & -\Omega_{\rm R}^*\\
\mathcal{A}_\downarrow^* & \mathcal{A}_{\downarrow*} & \Omega_{\rm R}^*& 0 & (\mathcal{H}_{0\downarrow}-\mu)/\hbar & 0 \\
-\mathcal{A}_{\downarrow*}^* & -\mathcal{A}_{\downarrow} & 0 & -\Omega_{\rm R} & 0 & -(\mathcal{H}_{0\downarrow}-\mu)/\hbar
\end{pmatrix},
\end{align}
where $\mathcal{H}_{0\uparrow}\equiv-\hbar^2\boldsymbol\nabla^2/2M+\hbar\delta+V_\uparrow$ and
$\mathcal{H}_{0\downarrow}\equiv-\hbar^2\boldsymbol\nabla^2/2M+V_\downarrow$.

In general, the Bogoliubov equation~\eqref{eqSM:Bog-eq} can be only solved numerically to yield the spectrum $\omega$ of collective excitations of the system. That said, for the uniform spin-polarized FM state as the normal state (below the self-ordering threshold) of $R=0$ (i.e., the running-wave pumps), it is possible to diagonalize the Bogoliubov matrix~\eqref{eqSM:Bog-matrix} analytically as we show in the following subsection. 

\subsection{The Threshold Pump Strength}

Let use now analyze the stability of the uniform spin-polarized FM normal state: $\alpha=0$ and $\psi_\tau=\sqrt{nf_\tau}$, where $n=N/V$ is the total density and $f_\tau=N_\tau/N$ is the fraction of the atoms in spin state $\tau$ with the normalization condition $f_\uparrow+f_\downarrow=1$. The coefficients~\eqref{eqSM:A-coeff} then simplify to
\begin{align} \label{eqSM:A-coeff-simple}
\mathcal{A}_\uparrow&=\mathcal{A}_{\uparrow*}^*=\eta_0\sqrt{nf_\downarrow}e^{ik_cy}\cos k_cx,\nonumber\\
\mathcal{A}_\downarrow&=\mathcal{A}_{\downarrow*}^*=\eta_0\sqrt{nf_\uparrow}e^{-ik_cy}\cos k_cx.
\end{align}
Due to the form of these coefficients, atomic fluctuations of the form
$\delta\hat{\psi}_{\tau\pm}(\mathbf{r})=[\delta\hat{\psi}_{\tau\pm}^{(+)}e^{ik_cy}+\delta\hat{\psi}_{\tau\pm}^{(-)}e^{-ik_cy}]\cos k_cx$ couple dominantly to the photonic fluctuations. Restricting to the subspace of these fluctuations, the Bogoliubov matrix takes the form,
\begin{align} \label{eq:M-matrix-restricted}
\mathbf{M}_{\rm B}=
\begin{pmatrix}
-\tilde\Delta_{0c} & 0 & 
0 & \sqrt{nf_\downarrow}\eta_0V/2 & \sqrt{nf_\uparrow}\eta_0V/2 & 0 &
\sqrt{nf_\downarrow}\eta_0V/2 & 0 & 0 & \sqrt{nf_\uparrow}\eta_0V/2 \\
0 & \tilde\Delta_{0c}^* & 
0 & -\sqrt{nf_\downarrow}\eta_0V/2 & -\sqrt{nf_\uparrow}\eta_0V/2 & 0 &
-\sqrt{nf_\downarrow}\eta_0V/2 & 0 & 0 & -\sqrt{nf_\uparrow}\eta_0V/2 \\
0  & 0  & \epsilon_{\uparrow} & 0 & 0 & 0 & 0 & 0 & 0 & 0 \\
-\sqrt{nf_\downarrow}\eta_0 & -\sqrt{nf_\downarrow}\eta_0  & 0 & -\epsilon_{\uparrow} & 0 & 0 & 0 & 0 & 0 & 0 \\
\sqrt{nf_\uparrow}\eta_0 & \sqrt{nf_\uparrow}\eta_0  & 0 & 0 & \epsilon_{\downarrow} & 0 & 0 & 0 & 0 & 0 \\
0 & 0  & 0 & 0 & 0 & -\epsilon_{\downarrow} & 0 & 0 & 0 & 0 \\
\sqrt{nf_\downarrow}\eta_0 & \sqrt{nf_\downarrow}\eta_0  & 0 & 0 & 0 & 0 & \epsilon_{\uparrow} & 0 & 0 & 0 \\
0  & 0  & 0 & 0 & 0 & 0 & 0 & -\epsilon_{\uparrow} & 0 & 0 \\
0  & 0  & 0 & 0 & 0 & 0 & 0 & 0 & \epsilon_{\downarrow} & 0 \\
-\sqrt{nf_\uparrow}\eta_0 & -\sqrt{nf_\uparrow}\eta_0  & 0 & 0 & 0 & 0 & 0 & 0 & 0 & -\epsilon_{\downarrow}
\end{pmatrix}.
\end{align}
where $\epsilon_{\uparrow}=2\omega_r+\delta$ and $\epsilon_{\downarrow}=2\omega_r$ for $\delta>0$ (with the chemical potential $\mu=\hbar\Omega_{01}^2/\Delta_1=\hbar\Omega_{02}^2/\Delta_2$), and $\epsilon_{\uparrow}=2\omega_r$ and $\epsilon_{\downarrow}=2\omega_r-\delta$ for $\delta<0$ (with the chemical potential $\mu=\hbar\Omega_{01}^2/\Delta_1+\hbar\delta=\hbar\Omega_{02}^2/\Delta_2+\hbar\delta$).
\begin{figure}[!b]
\centering
\includegraphics [width=0.87\textwidth]{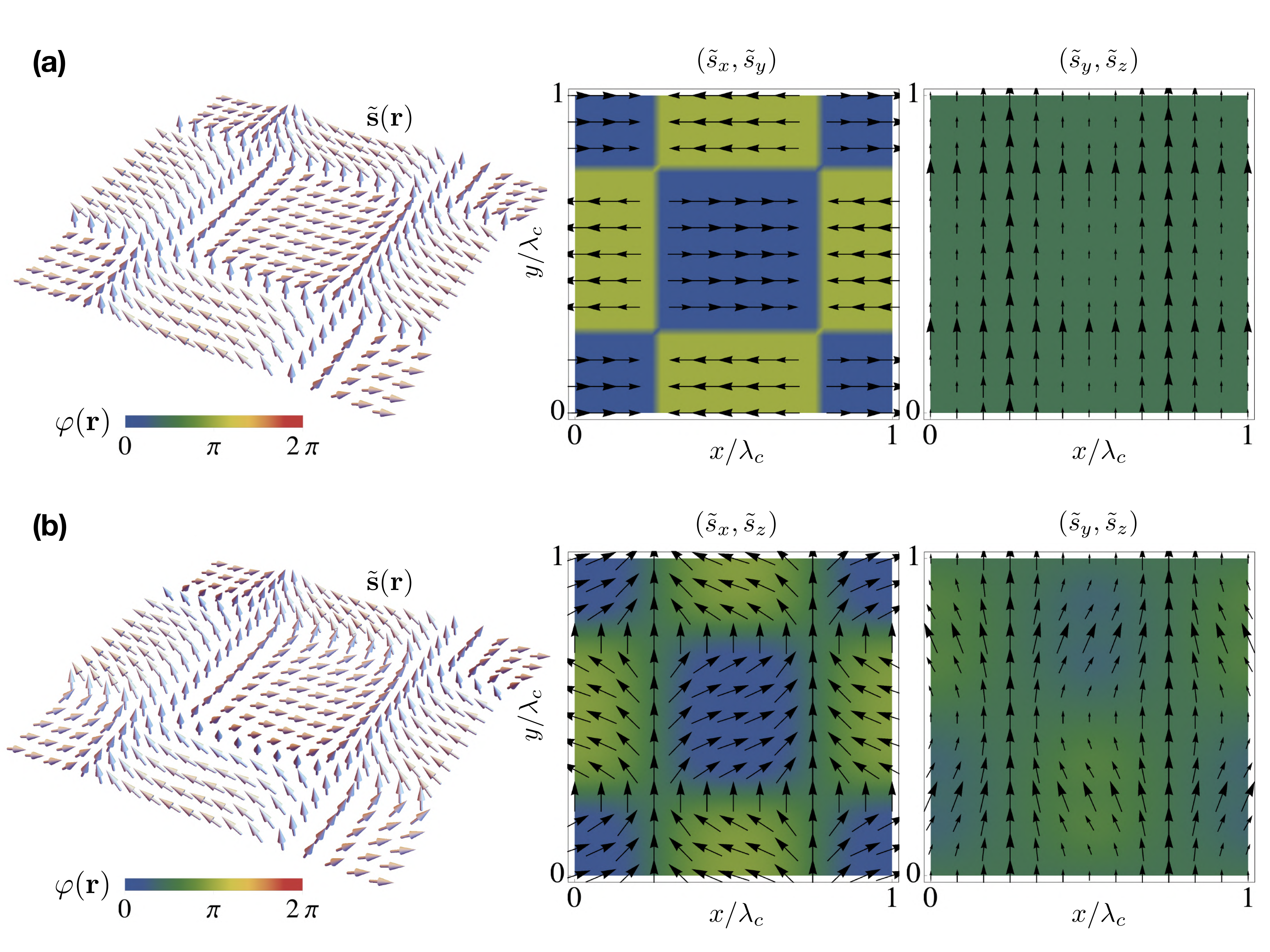}
\caption{The spin textures in the domain-wall AFM (a) and domain-wall SS (b) phases, which we did not present in the main text.
The parameters are the same as Figs.~2 and 3 in the main text.} 
\label{figSM:spin_texture_extra}
\end{figure}
Note that in addition to the balanced Raman condition $\eta_0\equiv\mathscr{G}_0\Omega_{01}/\Delta_1=\mathscr{G}_0\Omega_{02}/\Delta_2$, throughout this paper we have assumed for the sake of simplicity that $\Delta_1=\Delta_2$, which from the balanced Raman condition then follows $\Omega_{01}=\Omega_{02}$ .
The eigenvalues $\omega$ of $\mathbf{M}_{\rm B}$ is obtained via the 
tenth order characteristic equation $\text{Det}(\mathbf{M}_{\rm B}-\omega I_{10\times 10})=0$.
The zero frequency solution $\omega=0$ yields the self-ordering threshold for the spin-spiral phase, 
\begin{align}
\sqrt{N}\eta_{\rm SS}^{\rm c}=\pm\sqrt{-\frac{|\tilde\Delta_{0c}|^2}{\Re(\tilde\Delta_{0c})}
\left(1+\frac{|\delta|}{2\omega_r}\right)\omega_r},
\end{align}
where we have made use of the fact that $f_\uparrow=0$ for $\delta>0$ and $f_\downarrow=0$ for $\delta<0$.
For the parameters used in Figs.~2 and 3 in the main text, this yields a threshold value $\sqrt{N}\eta_{\rm SS}^{\rm c}\approx\pm3.57\omega_r$ for the SS phase which is in a good agreement with the mean-field numerical results.

\section{Some Additional Spin Textures}

Here we also illustrate the normalized local spin $\tilde{\mathbf s}(\mathbf{r})$ 
and its projections in both domain-wall AFM and domain-wall SS phases which we did not present in the main text. 
Figure~\ref{figSM:spin_texture_extra}(a) shows the spin textures on the domain-wall AFM phase and 
Fig.~\ref{figSM:spin_texture_extra}(b) the spin textures on the domain-wall SS phase.

\section{Details of the Numerical Approach}

We consider the mean-field limit of Eqs.~\eqref{eqSM:eff-Heisenberg-eqs-motion} and \eqref{eqSM:ss-a},
where we replace the photonic and atomic field operators with their 
corresponding quantum averages:
$\hat{a}_{\rm ss}\to\langle\hat{a}_{\rm ss}\rangle\equiv\alpha_{\rm ss}$ and 
$\hat{\Psi}\to\langle\hat{\Psi}\rangle\equiv\Psi=(\psi_\uparrow,\psi_\downarrow)^\top$.
We then look for self-consistent solutions of these mean-field equations. That is,
we start with a random value for $\alpha_{\rm ss}$ and then solve the two Schr\"{o}dinger 
equations for the spinor condensate wavefunctions $\Psi(\mathbf{r})$ in the position space. 
Using these calculated condensate wavefunctions,
we compute the new steady-sate field amplitude $\alpha_{\rm ss}$ via Eq.~\eqref{eqSM:ss-a}. This
procedure continues until the difference in successive values of $\alpha_{\rm ss}$ is less than a
pre-determined tolerance.

\end{document}